\providecommand{\lay}{{\rm lay}} 
\providecommand{\dist}{{\rm dist}}
\providecommand{\reg}{{\rm reg}} 

\documentclass[epj,final]{svjour}    
\usepackage{amssymb}
\usepackage{graphicx}
\begin{document}
%
\title{Stratifications of cellular patterns}
\author{C. Oguey\inst{1} \and N. Rivier\inst{2} \and T. Aste\inst{3} %
}                     
\institute{LPTM, Univ. de Cergy-Pontoise, 95031 Cergy-Pontoise, France,
\quad\emph{e-mail}: oguey@ptm.u-cergy.fr \and
LDFC, ULP, 67084 Strasbourg, France\and
Dep. of  Applied Mathematics, RSPHYSSE, Australian National University,
ACT 0200 Canberra, Australia}
\date{14 Nov 2002}                          
%
%
\abstract{
Geometrically, foams or covalent graphs can be decomposed into 
successive layers or strata.
Disorder of the underlying structure imposes a characteristic roughening of the layers.
Our main results are hysteresis and convergence in the layer sequences.\\
1) If the direction of construction is reversed,
the layers are different in the up and down sequences (irreversibility);
nevertheless, under suitable but non-restrictive conditions, the layers
come back, exactly, to the initial profile, a hysteresis phenomenon.\\
2) Layer sequences based on different initial conditions
(e.g. different starting cells)
converge, at least in the cylindrical geometry.
Jogs in layers may be represented as pairs of opposite dislocations,
moving erratically due to the disorder of the underlying structure
and ending up annihilating when colliding.
} 
\maketitle
%
\setlength{\unitlength}{1cm}

\section{Introduction} \label{intro}

Despite the broad range of their different material realisations
(liquid froths, metallic microstructures, polymeric foams, etc.),
foams and analogous disordered structures
have strong structural
similarities \cite{weaireRiv,weaireHutz}.
Even if universality has not been firmly
established for these structures,
many features obviously
do not depend on the details of the constituents, of the force fields, etc.
nor on the precise values of metric quantities.
Whence, our choice of describing foams and random patterns at the level
of topology.

To account for correlations and statistics beyond the one body properties,
it is natural to define configurations in
terms of topological distance.
Indeed, in foams, nearest neighbour cells are clearly defined
by sharing an interface. In the dual network, the cells are represented
by points connected by bonds, one for each facet in real space.
Thus the dual is a graph, which is sufficient to define topological distance
\cite{Aste,One,Dubertret,Rivier98}.

Covalent structures, like those occurring in glassy materials, are also
described in terms of graphs: the atoms, molecules or clusters sit
at the nodes and the covalent links define the bonds.
Since covalent interactions are carried by quantum
electron clouds, defining the bonds is not always
free from ambiguities. This is still more so when non
covalent interactions are involved. In these cases, geometric constructions,
such as the Dirichlet-Voronoi one, may complement or replace
physico-chemical criteria. 

The minimal number of (dual, in the case of foams) bonds, or steps, needed to join two nodes
defines a {\em topological distance} (or simply distance, since no other notion
of distance will be considered here).

A layer is the set of nodes / cells at a fixed distance $j$ from an origin $O$.
Partitioning the entire foam into successive layers $j=1,2,3,\ldots$
provides a stratification of the cellular pattern.

There are many reasons for improving our understanding of stratifications.
The number of nodes / cells, in successive layers ---the population, for short---
gives almost the same
information as the pair correlation function \cite{One,Rivier98,OgRiv}.
As is well known, the correlations are related to the response of the system to
all kinds of solicitations. In disordered materials, this question is
of particular interest: is the response coded in the geometry and how ?
Reciprocally, beyond elasticity, external actions may
modify the structure. How ?
Aging is a common characteristic of glassy materials which almost never reach
equilibrium; aging may occur spontaneously or under external influence,
often in an inhomogeneous way.

All these questions involve structure. 
Our purpose, here, is to analyse some of the fundamental geometric tools,
to set the ground for further research.
Ultimately, energy should be considered.
But, in complex systems, the step from geometry to energy
is often easier than understanding the geometry.
Foams are paradigmatic in this respect:
to first approximation, energy is film length (in 2D) or
area (in 3D) times a constant (surface tension).

Viewed as dynamical processes, considering $j$ as time, layer sequences
represent the successive stages of signals, fronts, epidemics propagating
at unit velocity.
There is a close analogy with aggregation-deposition and related problems
(cf. \cite{OgRiv} and refs. therein).

One of the differences, however, is that
the underlying foam is given {\it a priori} in its full integrity.
The stratification is a supplementary structure ---an ordered partition--- of the foam.
It is therefore necessary to disentangle what is general to layers,
from what depends specifically on the underlying structure. 
Notably, the same structure can have many possible stratifications.
A preliminary answer to the question of sorting these stratifications
will be found here through convergence.

The arbitrariness comes from the choice of the origin.
It corresponds to the large collection of different stratifications that
are possible for partitioning a given foam or covalent structure.
What symmetry covers these equivalent choices ?
For example, in \cite{One}, the leading asymptotic behaviour of the
layer population $K_j$ was
found (numerically) to be independent of the central cell (the origin).
Here, we give a strong support to this hypothesis
by showing that the stratifications do actually converge.
The central cell may even be replaced, as the origin, by a central set of cells.

Another distinction is of importance:
the elementary models of aggregation, such as the Eden model, are random processes
on regular lattices \cite{Godreche}. In our case, the underlying structure
---a foam--- is random, with quenched disorder, whereas the process ---stratifying---
obeys a fixed, deterministic, rule (without any randomness).
Some geometrical features, such as roughness, are similar in both types of systems
\cite{OgRiv}. In the present paper, we insist on aspects which are more specific
to the second class: (ir)re\-versibility, hysteresis, reciprocity, convergence.

\subsection{Layers}

\begin{figure*}
\center
\begin{tabular}{lll}
\hspace{-3mm}\includegraphics[width=6cm,height=12cm]{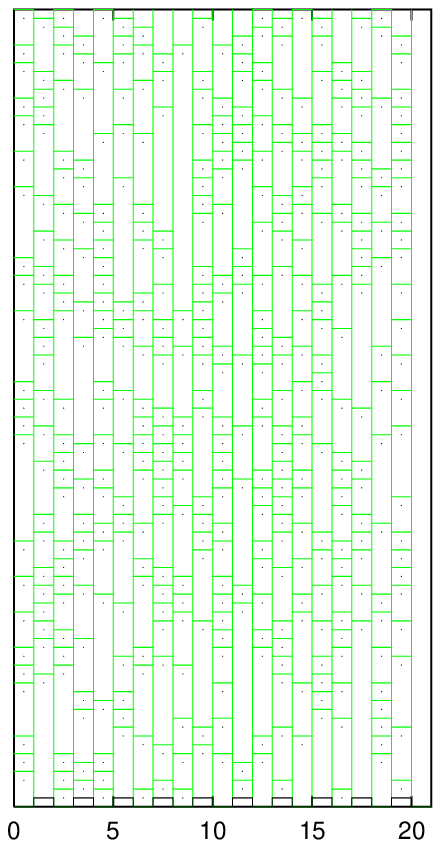}&
\hspace{-6mm}\includegraphics[width=6cm,height=12cm]{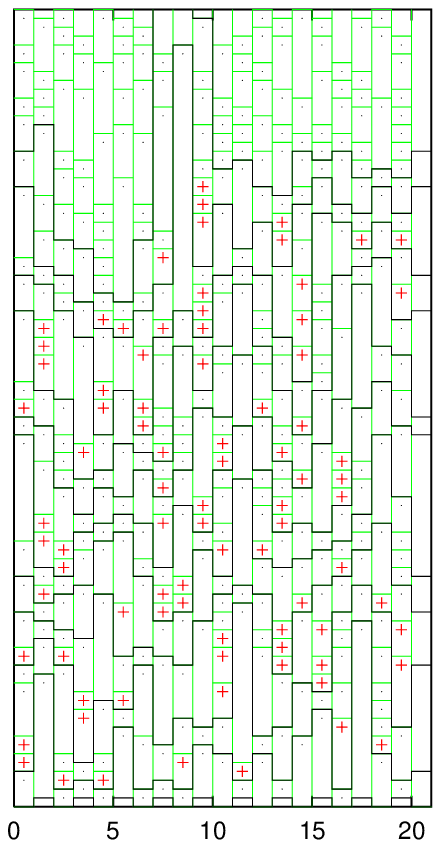}&
\hspace{-6mm}\includegraphics[width=6cm,height=12cm]{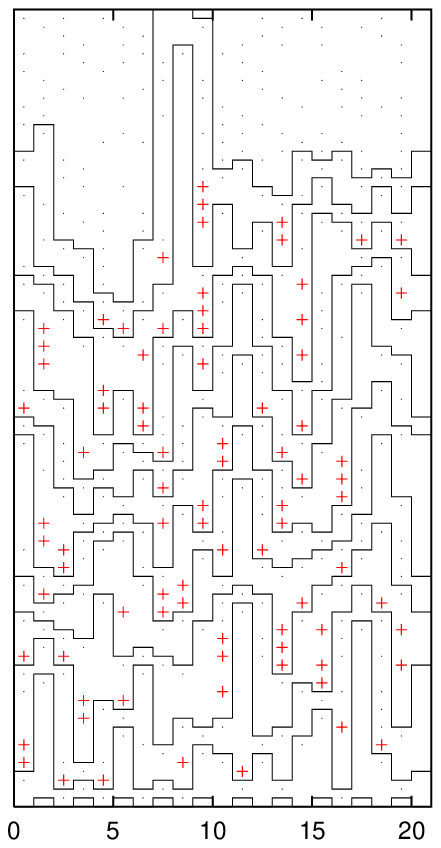}\\
Bare cells of the columnar foam; top&
First layers superposed on the cells;&
Layer boundaries only;\\
square of each cell marked by a dot.&the defects are now marked by '+'.
&the regular cells remain dots '$\cdot$'.\\
\end{tabular}
\caption{\label{didacticol}
Graphical conventions for displaying cells, defects and layers,
shown on an example with $L=20$.}
\end{figure*}

In the foam, we classify the cells in terms of topological distance:
layer number $j$ (layer $j$ or lay($j$) for short) is the set of cells at
distance $j$ from $O$.
The origin $O$ may be a single cell or a cluster of
cells (concentric geometry) or a connected set such as a row of cells (going
once around the cylinder in cylindrical geometry, or infinitely long
in open Euclidean geometry).

(In general, the dual of a cellular structure is a multi-graph.
For almost all natural foams, it is a simple graph, that is, neighbouring cells
have at most one facet in common.
This technical assumption is not essential, and it could easily be lifted
if necessary, but it makes the presentation simpler.)

For cellular structures, this defines layers through the dual.
It is also possible to operate in the direct cellular network as follows
\cite{Rivier98,fortesPina}.
Consider an initial cluster $O$, called the origin and labelled $j=0$;
the cells in contact with $O$ constitute the first layer. 
Then, inductively for $j=1, 2,\ldots$, layer $j$ is made
of all the cells, not yet counted, which are in contact with layer $j-1$.

If, as in modelling chemical structures, the origin is
a single vertex ---an atom in the compound---, 
then the layers are coordination shells \cite{OKeeffe,Conway,Baake}.
So layers, coronas \cite{Deza} and coordination shells are synonyms.
We also name them strata, because they partition the foam
---the set of vertices in the dual--- into an ordered collection of subsets,
making altogether a {\em stratification} (or foliation).

The embedding space is either Euclidean (the plane in 2D)
or a cylinder equivalent to a domain of bounded base
with periodic boundary conditions
in the $x$  direction (a circle in 2D) and infinite along the axis
of the cylinder ($y$ coordinate).

In layer $j$,
\vspace{-3mm}
\begin{itemize}
\item every cell is neighbour of at least one cell in lay$(j-1)$;
\item some cells, called {\em regular}, are also neighbour of cells in lay$(j+1)$;
\item the other cells, not sharing any edge (facet in 3D) with cells in
lay$(j+1)$, are called {\em defects}.
\end{itemize}

The first statement above is part of the definition of layer $j$;
the next two are definitions of regular and defect cells in the layer.
As we shall see, defects are sources of frustration, curvature and non-triviality
of the stratification.

In summary, a stratification
$\ell=\{\ell_j\}_{j\geq 0}$
is a partition of the foam
(or of the set of nodes in covalent graphs) into layers --- the strata
$\ell_j,\ j= 0,1,2,\ldots$.
Each layer is the set of cells at distance $j$ from $O$:
$\ell_j = \{c\; |\; \dist(c,O)=j\}$.

{\em Shell} $h_j$ is defined as the outer boundary of layer $j$;
it is the contour lying between layers $j$ and $j+1$.

\subsection{The columnar model} \label{columns}

This toy model (the columns) is a useful laboratory
for the structure of foams and as a model of growth.
It is a lattice version of the Poisson partition of Fortes \cite{fortes}.
We use it for illustration but most of the features presented here
are valid generally, not limited to this example.

The relevance and limits of this model were discussed in
\cite{OgRiv}.
For completeness' sake, we recall the basics here.
 
The model is a 2D packing of columnar cells, each of width 1 (in the
horizontal, $x$, direction) and of random length $s$ (height in the vertical,
$y$, direction).
The sizes ($s$ is both length and area) of the individual cells are taken as
independent random even numbers, identically distributed with exponential law :
\begin{equation}    \label{celldistr}
  \Pr(s) = \frac{1-z}{z}\; z^{s/2}, \ s = 2,4,6\ldots.
\end{equation}
The parameter $z$ has a fixed value in $]0,1[$.
It controls the mean cell size through $\langle s\rangle = 2/(1-z)$.
Unless otherwise stated, we will take $z=1/2$, $\langle s\rangle = 4$.
 
The foam lies on a semi-infinite vertical cylinder,
meaning that it is periodic, with period $L$, in the $x$ direction. 
The height $s$ of each cell is an {\em even} random number.
With ground $y_0(x)=x$ mod 2 (crenellated profile),
this ensures that the vertices
have coordination 3, as in real foams.
The system is unbounded in the positive $y$ direction.

To avoid overloading the pictures, the graphical convention
of Figure \ref{didacticol} (right) will be used:
the cell boundaries are not drawn; the lines are layer boundaries 
(= shells $h_j, j=0,1,2,3,\ldots$).
The top square of each cell is marked by a
dot (.) when the cell is regular, by a cross (+) when it is a defect
\cite{Aste,Rivier98,OgRiv}.

\section{Up and down: irreversibility} \label{irreversibility}

We compare different stratifications on the {\em same} cellular pattern.
In this section, we build two sets of layers,
one with distance increasing upwards (stratification from the bottom up),
the other with distance increasing
downwards (stratification from an origin at the top).
Later, in Section \ref{convergence},
we compare stratifications rising in the same direction but based on
different origins (grounds). The question is whether they match, and, if
so, how ?

\subsection{Up and down}  \label{updown}

\begin{figure*}
\center
\begin{tabular}{lll}
\hspace{-.6cm}\includegraphics[width=.36\textwidth,height=13cm]{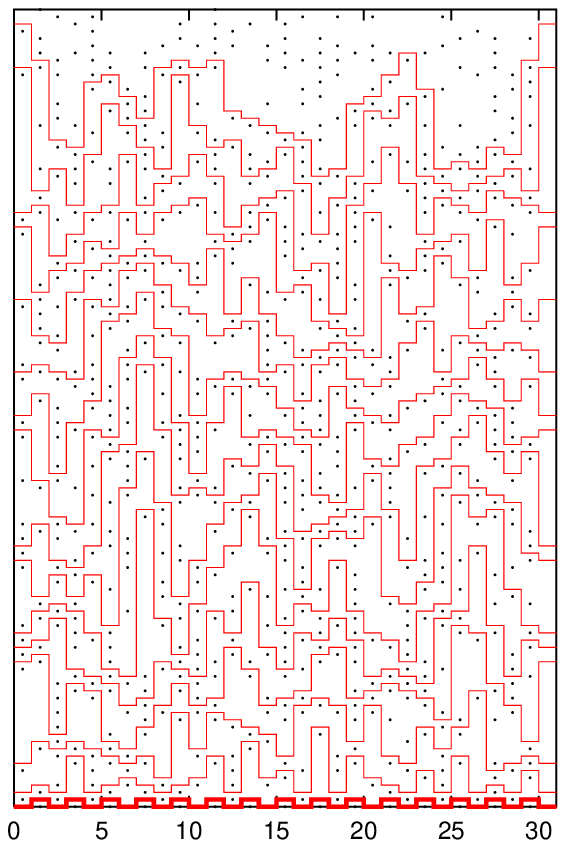}&%
\hspace{-1cm}\includegraphics[width=.36\textwidth,height=13cm]{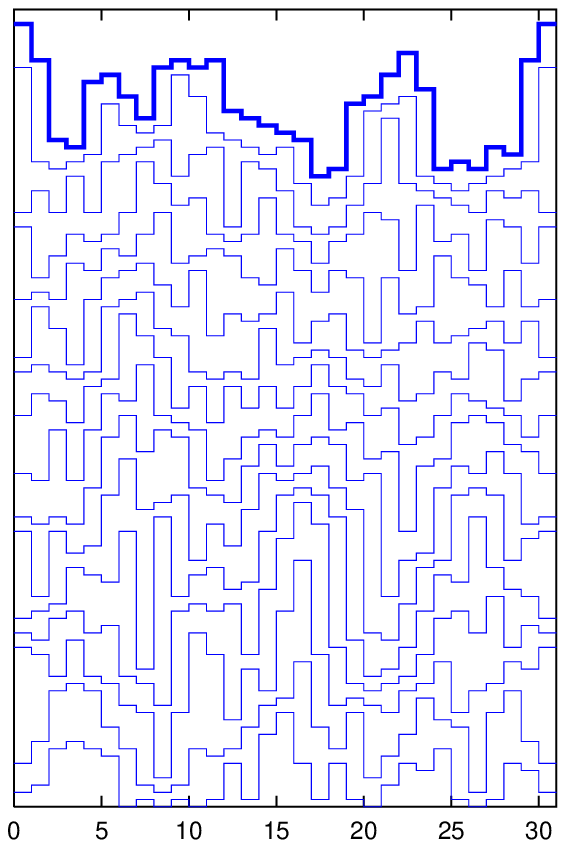}&%
\hspace{-1cm}\includegraphics[width=.36\textwidth,height=13cm]{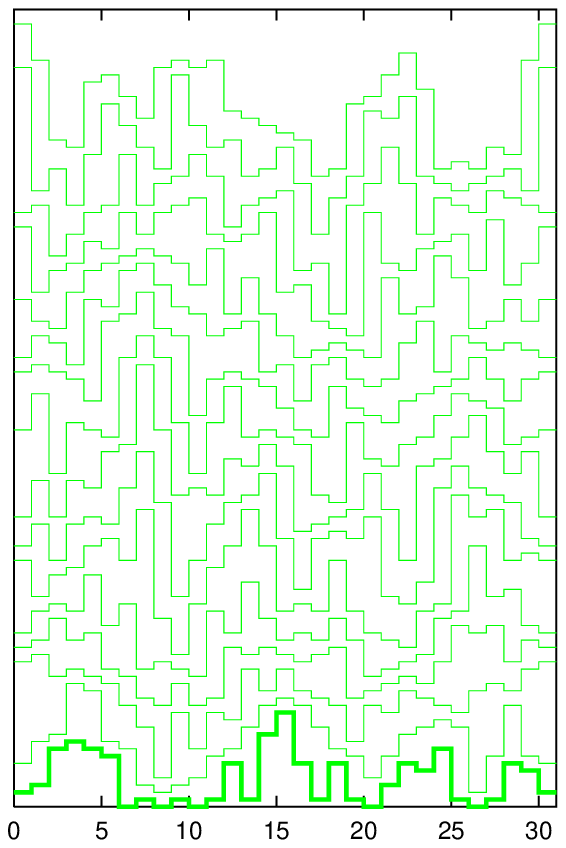}\\
Layers $j=0, \ldots, 15$, up.&
Idem, descending.&
Idem, up again.\\
\end{tabular}
\caption{\label{updownfig}
$L=30$ :  stratifications of 15 layers up, down and up again.
The ground $j=0$ is highlighted in each case. The profiles $h_{j}(x)$
of the two upwards stratifications are identical for $j \geq 12$.}
\end{figure*}

Starting from an origin $A_0=O$ or a ground $h_0$, if we build the layers upwards
$A=\{A_1,A_2,\ldots,A_j,\ldots\}$,
stop at some
$j=d$ and then, taking
layer $A_{d}$ as a new origin $O'$
(equivalent to setting shell $h_{d-1}$ as a starting profile $h'_0$),
build new layers $A'=\{A'_0=O',A'_1,A'_2,\ldots\}$ downwards, 
this new stratification $A'$ does {\em not}
coincide with the former 
(even if we compare just the regular parts).
The top most layer of $A$ coincides with the origin of $A'$, by construction. 
But then, some cells, qualified as defects in upward layers
switch to being regular in downward layers
and vice versa, etc. 
Since these switches cumulate during buildup of the stratification, we
may expect that the coherence between the two reverse stratifications $A$
and $A'$ is rapidly lost. This appears to be the case, at first sight (see
Figure \ref{updownfig} left and middle).

Notably, the last shell going down, $h'_{d-1}$, is different
from the upward starting ground $h_0$. It lies in a neighbourhood of $h_0$,
but it is different. This difference will be described in Section \ref{geometry}.

\subsection{Back up}

What happens if we go back again ?
Take the last down shell $h'_{d-1}$ as a new ground,
$h^B_0$, and build another stratification
$B=\{B_0=A'_{d},B_1,B_2,\ldots\}$ climbing up again.
This third stratification $B=\{B_j\}$ is different from the first two: 
different from $A'$ because of irreversibility;
different from $A$ because the new ground, $h^B_0$, is not, in general, a shell
$h_j$ of the first stratification (Figure \ref{updownfig} right).

Nevertheless, after climbing up, building $B$ up to $B_{d}$,
the last profile fits exactly the same profile as the first crest:
$h^B_{d-1} = h_{d-1}$, $B_{d}=A_{d}$.
This will be proved later.

Further up and down processes repeat $A'$ and $B$.
Indeed, the next stratification $B'$ (downwards) is degenerate with $A'$ since
it starts from the same origin, and so on.
Recall that all these stratifications are based on the same, fixed, but random, foam.

\subsection{Hysteresis}

We have therefore a {\em hysteresis} cycle,
caused by the presence of defects.
Indeed, defect-free stratifications are reversible.
Examples of these are the rows and columns parallel
to the square basis in le Caer's construction \cite{leCaer,schliecker},
the vertical columns in the columnar model, or even the horizontal
layers $\{\ell^0_j\}$ after defect coalescence.

All these are flat or pure gauge models
like the Mattis model for spin glasses \cite{frustration}.
But, let us stress this point, these models admitting defect-free
stratifications are not generic.
Notably, in le Caer's model, there are special correlations between neighbouring
cells \cite{schliecker}.
Topologically random foams are not flat.
Hysteresis might even be taken as a measure of non trivial disorder.

\section{The geometry of layers} \label{geometry}

In this section, we present the stratifications as analogous
to foliated structures. In particular, a proof is given of the fact
that the extreme layers are exactly recovered by the down-up procedure. 

\subsection{Layers as sets} \label{secsets}

The distance between sets $A, B$ is defined as
\begin{equation} \label{aqsets}
\dist(A,B) = \min\{\dist(a,b) | a \in A, b \in B\}.
\end{equation} 
In this sense, layer $j$, as a set of cells, is at distance $j$ from
the origin $O$ which can consist of more than a single cell.
In fact, by definition,
{\em all} the cells of $\lay(j)$ are at
distance $j$ from $O$. But the converse is not true: {\em not} all
the cells of $O$ are at minimal distance $j$ from 
$\lay(j)$. This is a first indication of irreversibility.

\subsection{Geodesic sections}

Let us go on.
By definition, any cell $c_j$ of lay$(j)$ has at least one neighbour
$c_{j-1}$ in lay$(j-1)$;
$c_{j-1}$ has a neighbour $c_{j-2}$ in lay$(j-2)$ etc. down to 
some {\em root} cell $c_0$ in $O$.
Thus, there is always at least one {\em connected section}
linking any cell $c_j$ in layer $j$ to $O$.
In the dual, these connected sections are lines of minimal length
$j=\dist(c_j,O)$, i.e. topological {\em geodesics}, linking $O$ and lay$(j)$.
Moreover, these sections consist of regular cells exclusively.
In particular, defects in $O$ cannot be root cells.

Stratification is analogous to foliation in differential geometry.
The layers are the leaves, and any
section can serve as base space (isomorphic $\mathbb{Z}$ or some
subinterval of $\mathbb{Z}$).

The layer structure is robust along these sections:
there is exactly one cell per layer crossed.
Along these lines, each step is a move from a layer to the next one
--- upward or downward. 
Thus, the sequence of layer numbers $j=1,2,\ldots$ coincides with topological
distance along these lines (counted, respectively, from bottom up,
or from top down ). The set of linking geodesics constitutes an orthogonal
{\em skeleton} for the stratification.

As already stated, there is a section linked to every cell in lay$(j)$,
but not every cell $o$ of $O$ is at 
distance $j$ from lay$(j)$; only the root cells are.
Linking geodesics starting from different top cells may fuse
on the way down. So the whole set (of linking geodesics)
is a forest with branches attached to every cell
of lay$(j)$ but only a few root cells in lay$(0)=O$.
(Note that there may be more than one geodesic
connecting two given cells).
Another forest pattern was introduced in \cite{Rivier98}.

The space left ---that is, the part of the foam not covered by the skeleton---
is the place where irreversibility occurs; the layers down differ from the layers up.

\subsection{Up and down revisited: parallel layers} \label{paralellay}

Two sets $A$ and $B$ are {\em parallel} if there is a positive
number $d$ such that 
i) all the $a\in A$ are at the same distance $d$ to $B$ and 
ii) all $b\in B$ are at the same distance $d$ to A.

With respect to stratifications, where the sets are sets of cells and
distance is topological distance,
parallel sets enjoy special properties.
If $A$ and $B$ are parallel at distance $d$:
\begin{itemize}
\item In the stratification based on $A$, $B$ is a subset of the $d$th layer:
lay$(0)=A \Rightarrow B \subset {\rm lay}(d)$.
\item All the cells of $A$ are root cells.
\item Conversely, $A$ is a subset of the $d$th layer based on $B$:
lay$(0)=B \Rightarrow A \subset {\rm lay}(d)$.
\item All the cells of $B$ are root cells in this stratification. 
\end{itemize}

In the notations of Section \ref{irreversibility},
we now show that the layers $A_d$ and $A'_d$ are parallel at distance $d$.
The proof requires a few basic (in)equalities.

\subsubsection*{Up $\uparrow$}
In the up stratification $A_0=O, .., A_j, ..$
---as in any stra\-tif\-ication---
dist$(\lay(j), O) = j$ only implies 
dist$(\lay(j), o) \geq j$ for an arbitrary cell $o$ of $O$.
Equality holds if and only if $o$ is a root cell $c_0$ for some geodesic section.
Moreover, equality must hold for at least one cell;
there always is at least one root cell in $O$.

\subsubsection*{Down  $\downarrow$}
Consider the down stratification $A'_0,A'_1, A'_2, \ldots,A'_d$.
$A'_0$ consists of all the cells of $A_d$ of the up stratification.
Call $\{b_i\}$ the cells of layer $A'_d$. $\{b_i\}$ includes all
the root cells of $O$.
All the others must lye below $\{b_i\}$ because they
satisfy strict inequality: dist$(o, A_d) > d$.

\subsubsection*{Parallelism $\updownarrow$}
Thus, $A'_0$ and $A'_d$ are parallel.

Indeed, i) dist$(b, A'_0)=d,\;\forall\, b\in A'_d$ holds by definition of layer $A'_d$.
To see that ii) dist$(a,A'_d)=d,\;\forall\, a\in A'_0$, note that
the construction of the layers implies dist$(a,A'_d)\geq d$.
On the other hand, in the up stratification, there is a cell $c_0 \in O$ at 
dist$(a,c_0)=d$; being root, $c_0$ also belongs to $A'_d$.
Therefore dist$(a,A'_d)\leq d$, which proves equality ii).

\subsubsection*{Remarks}

\begin{enumerate}
\item
Note that $A'_0$, which was set equal to $A_d$,
contains no defect for the downward stratification.
Indeed, any cell $c\in A_d$ is at distance 1 of a (regular) cell of $A_{d-1}$
and any regular cell of $A_{d-1}$ is reached this way, implying
$A^{\reg}_{d-1}\subset A'_1$. Now a defect $c_d$ in $A'_0$ would be at distance
at least 2 from $A'_1$, in contradiction with
$c_d\in A_d=\{c\,|\,\dist(c,A^{\reg}_{d-1})=1\}$.
\item
Going up and down establishes a 'reciprocity' relation between
layers $A'_0 = A_d$ and $A'_d$, slightly stronger than parallelism.
Such a reciprocity does not hold in general;
most often, two layers in the same sequence are not even parallel.
In order to get a pair of reciprocal layers, a precise procedure
must be followed, such as the up-down trick.
\item
Apart from that, nothing special is assumed
on either the foam or the original layer. 
The point where we turn back ($j=d$) is chosen arbitrarily; layer $A_d$
is absolutely normal, with neither more nor less defects than any other.
Actually, if $d$ is large enough, the system basically forgets its
initial conditions. Another stratification, in the same sense but
starting from another origin, would generate the same layers at sufficiently
large $j$.
\item
A stack of successive layers has minimal thickness when it is delimited by
a pair of reciprocal layers.
With the previous notations, this means
dist$(o, A_d) \geq d = \dist(c, A_d)$ for all $o$ in $A_0$, $c$ in $A'_d$,
whatever $O$ we start from.
\item
Reciprocity does not mean reversibility of the process (layer sequence)
nor does it coincide with parallelism. It implies parallelism,
but parallelism is weaker because it
misses some completeness condition, as shown in the following example.
\end{enumerate}

\subsubsection*{An example: concentric stratifications}
Take a single cell $O=\{o\}$ as origin and build the concentric stratification
around it. Then $o$ must be a root cell.
Going up and down (out and in, implying a new stratification inwards)
brings one back to a cluster $C$ containing the starting cell $o$
possibly surrounded by other cells.
Lay$(j)$, which, in this case, is the topological circle of radius $j$, 
and its centre $\{o\}$ are parallel (according to our definition).
But they are not 'reciprocal'; up-down does not come back to only $\{o\}$.  
Cluster $C$, on the other hand, is both parallel to, and in
reciprocity relation with, the topological circle since it
was constructed so.

Notice that $o$ is parallel to any topological circle around it
(any azimuthal layer at distance $j=1,2,\ldots)$.
But cluster $C$ is parallel only to some specific circle(s),
where the turn back is done, or could be done, in order to
get $C$ exactly.

Irreversibility, or hysteresis, is the fact that, in between $C$ and the circle $j$,
the outwards and inwards stratifications are different; this is visible
only if $j>2$.

\section{Convergence of the stratifications, dependence on ground} \label{convergence}

The choice of the origin $O$ is arbitrary.
One may choose a single cell, and obtain concentric layers.
But choosing a horizontal ground is better adapted to cylindrical geometry.
Consider a definite foam on a cylinder.
Call $A=\{A_j\}_{j\geq 0}$ the stratification based on
$y= h^{A}_0(x)\simeq 0$ (a connected set of cell boundaries;
$y$ is the coordinate along the cylinder axis).

For the same foam, we could take as origin another
profile $\{h^B_0(x)\; |\; x=0,.., L-1\}$ following other cell edges.
Let $B=\{B_j\}_{j\geq 0}$ be the stratification based on $h^B_0$.
How do $A$ and $B$ compare ?

\subsubsection*{Global shift}
If $h^B_0$ is a shell of $A$, say $h^B_0=h^A_k$ for some integer $k$,
then, trivially,
$B_j=A_{j+k},\; \forall\, j>0$. The two layer sequences are identical;
only their label differ by an integer $k$ (an irrelevant phase shift).
Therefore, only profiles $h_0$ with centre of mass near $y=0$ need
be considered. 

\subsubsection*{Convergence}
From numerical simulations on columns and topological foams
with randomly generated $h_0^B$,
we observe that the stratifications $\{A_j\},\{B_j\}$ converge:
for any $h^B_0$, there are integers $J, k$ such that 
$B_j=A_{j+k}$ for all $j>J$.

The rate of convergence will be discussed later (Sec. \ref{convergenceRate}).
First, we analyse the phenomenon in terms of dislocations.

\subsection{Dislocation pairs in the stratifications} \label{podium}

Apart from the flat ground $h^A_0(x)=0$, the simplest starting ground
is a 'podium': $h^B_0(x) = 1$ for $x_+<x<x_-$, 
        $= 0$ otherwise (in vertical units of layers).
The steps at $x_+, x_-$ are a pair of dislocations in the stratification,
with strengths +1, -1.
Because of periodic boundary conditions, the strengths must sum up to $0$.

Let us compare $A$, based on a fixed ground, with a stratifications $B$,
based on a podium $h^B_0$ of width $w$
and of height 1 in units of $A$-layer thickness.
At $j=0$ the dislocations are at the ends of the podium:
$x_+(0)$ and $x_-(0)$ with $|x_+(0)- x_-(0)|=w$.
Choosing the maximal distance $w \simeq L/2$ will give an estimate
of the convergence time for more general situations.

At any later 'time' $j$,
away from the dislocations $x_\pm(j)$, the two layer systems
(profiles, inclusions, etc.) are the same, except for a shift
of 1 in numbering between the two dislocations.
The differences are confined to the region near $x_+$ and $x_-$
where the numbering makes steps.

\begin{figure} \centering
\includegraphics[width=7.5cm]{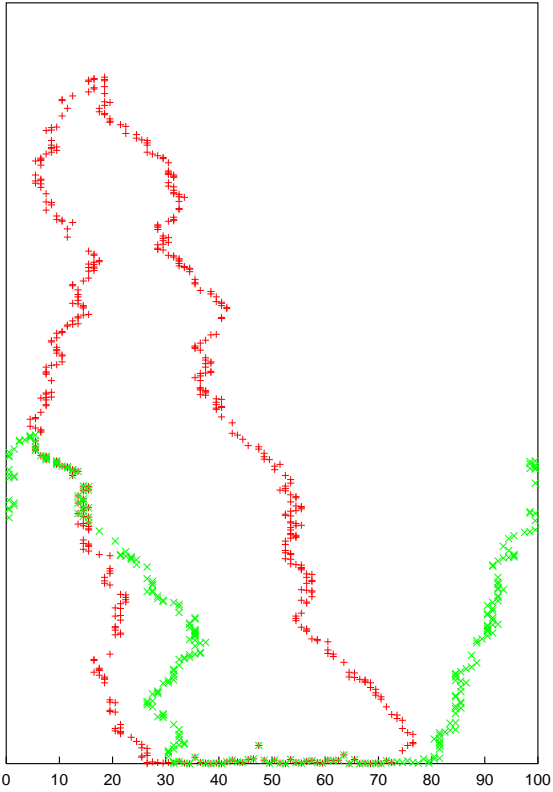}
\caption{\label{fdislocs} 
Two examples of layer convergence.
For a given, columnar foam, we compare two stratifications:
$B$ is based on a podium at $j=0$; the
reference $A$ is based on a flat ground.
Only the cells which are defect in one stratification ($A$ or $B$) but not in the
other are marked. The podium has width $w = L/2$, and centre at $L/2$ in one case (+)
and at $L/2+6$ in the other ($\times$).
In both cases, the pair of dislocations annihilates at some time $j = J$
(different in each case).
In the first case, the layers end up in phase.
In the second case, the final time shift is one (as if the podium had covered
a full layer).
The sample contains $(L=100) \times 400$ cells. }
\end{figure}

As can be seen in Figure \ref{fdislocs}, where are only marked cells which are
defects in one stratification but not in the other,
the differences look like two random walks which ultimately annihilate,
as in a 'diffusion-reaction' phenomenon.

The convergence occurs at time of first collision $J$, when the 
opposite dislocation meet for the first time and cancel.
The layers agree from there on
because, for a fixed underlying foam, the process
$$\ldots \rightarrow \lay(j-1) \rightarrow \lay(j) \rightarrow \lay(j+1) \rightarrow \ldots
$$
is deterministic.

Due to periodicity in the $x$ direction, the dislocations may fuse on one side
(with vanishing $h^B_j-h^A_j>0$ region),
or on the other (vanishing $h^B_j-h^A_j=0$ region).
Convergence means that $A_j=B_j$ in the former case, 
$B_j=A_{j+1}$ in the latter, for $j>J$.

Incidentally, in a crystalline foam, the analogous trajectories
would be periodic in space (ballistic regime).
Therefore convergence would occur in time $j=J$ linear in $L$
(or not at all, when the lines $x_+, x_-$ are parallel).

In random foams, convergence depends on disorder.
Here it is faster than in standard diffusion;
the spreading grows with time to the exponent $1/ z = 2/3$ instead of $1/2$.
(see Sec. \ref{convergenceRate}).

\begin{description}
\item[\em Remark] The various stratifications are made over a given, random
structure. Drawing the successive layers is therefore an entirely
{\em deterministic} process over the same random structure.
Convergence is like many of these mechanisms for finding successive key cards in a given,
shuffled pack. Once
two stratifications are in phase at some time $J$, they remain in phase
thereafter.
\end{description}

Because of periodic boundary conditions, a general ground $h_0$ can always be
decomposed into dislocation pairs (+/-1 steps).
When, initially, there is a large density of dislocations
(highly corrugated $h^B_0)$), many dislocation pairs cancel at small $j$
because the partners are initially close to each other;
this holds for random (diffusion) and crystalline (ballistic) foams.
The ultimate convergence of the stratifications
is controlled by the few dislocations that survive at longer time ($j$).
This is further analysed in Sec. \ref{convergenceRate}.

\subsection{Attractor}

Clearly, the outcome of the convergence
is a layer system ---a stratification--- which is a stable attractor.

Specifically, there are two stationary stratifications: one up and one down.

\subsection{Convergence rate} \label{convergenceRate}

\begin{figure}  
\centering \includegraphics[width=8.4cm]{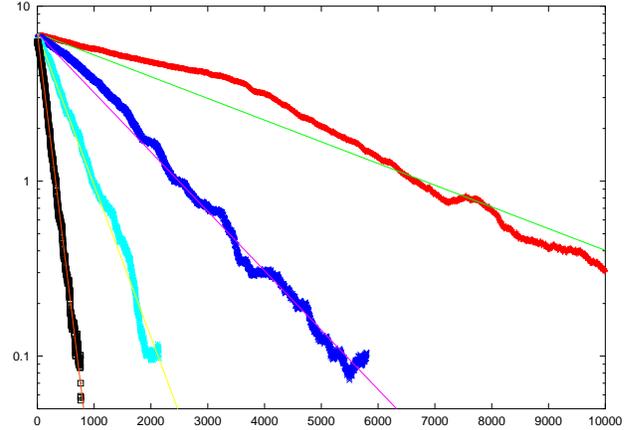}
\caption{\label{hconv} 
$\langle \Delta h_j\rangle = \langle \min_k |h_j^B - h_{j+k}^A| \rangle$ 
as a function of $j$ for $L=100, 200, 400, 800$.
(Average over 50 stratifications $B$ $\times$ 50 different foam structures).
Fits are of the form $c \exp(-j/\tau)$.
For each foam, the 50 stratifications $B$ have randomly chosen grounds.
Each stratification contains more than 10000 layers.}
\end{figure}

\begin{figure}   
\centering \includegraphics[width=8cm]{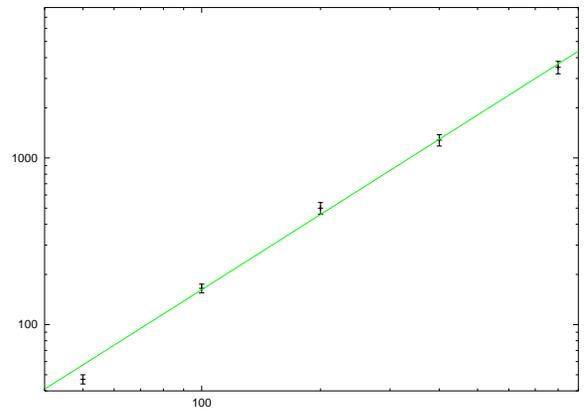}
\caption{\label{fitlog} 
Correlation time $\tau$ in function of $L$, log-log.
The line is $0.162\times L^{1.5}$. }
\end{figure}

If the random motion of the dislocations is governed by some cooperative
phenomena related to roughening,
then $J$ should be the time needed to reach 
$\xi_j = L$, $\xi$ being the correlation length along the layer.

First, at a fixed sample width $L$, convergence occurs at an exponential rate.
This has been checked by measuring the mean distance between the profiles,
$\langle \Delta h_j\rangle$, where $\Delta h_j = \min_k |h_j^B - h_{j+k}^A|$
(Figure \ref{hconv}).
Indeed, the correlation ``time'' $\tau$, defined by
$\langle \Delta h_j\rangle \propto \exp(-j/\tau)$, is finite as long as
the maximal possible distance between dislocations is bounded,
as it is for finite $L$.

The long time pseudo-diffusion process is manifest in the dependence
$\tau(L)$ on sample size $L$.
In the columnar model,
which has been shown to fall into the KPZ universality class \cite{OgRiv,KPZ},
the characteristic ``time'' $\tau \simeq \langle J\rangle$
scales as $\tau \sim L^z = L^{1.5}$; $z=1.5$ is the dynamic exponent.
This prediction is well confirmed by our simulations.
In the range of large $L$, the plot (Figure \ref{fitlog}) shows a scaling
behaviour fitting a power law $\tau \sim L^{1.5}$
in accordance with KPZ.

\section{Discussion, conclusions, perspectives...}

\subsection{Summary}

For foams or random covalent structures,
we have shown that the layer sequences are irreversible.
The stratifications in one direction and the other differ even if the
two sequences share a whole layer (this can be done by the up-down trick).
The hysteresis between up and down stratifications
is due to topological defects, inherently present when the disorder in non trivial.
The up-down procedure leads to topologically parallel layers at distance $d$,
enjoying special properties. 
\begin{itemize} 
\item Reciprocity: each one may be reached from the other
by building a sequence of $d$ layers.
\item Minimal thickness of the enclosed stack: any set of $d$ successive layers
ending at one of the parallel layers, but based on another initial condition
at $j=0$, will have a thickness larger than the strip bounded by the parallel
layers.
\end{itemize}

On a given fixed foam, the stratifications based on different origins converge to an
attractor, one for each of the two directions (in cylindrical geometry).
This pair of attractive stratification appears to be specific of the
underlying cellular pattern.
The characteristic time for convergence, $\tau \sim L^{1.5}$,
agrees with KPZ universality class,
as long as the probability distribution decays rapidly 
for cells with a large number of sides $n$
(exponentially, or as $n^{-\kappa}$ with $\kappa$ large enough) \cite{oguey}.
This has been confirmed by numerical simulations on the columnar model.

As a consequence, stratifications built on two foam samples differing only by local
perturbations ---topological transformations like neighbour exchanges,
cell birth or coalescence, etc.--- will also converge, even if
the convergence is, in practical respects, slow (see Sec. \ref{convergenceRate}).

As proven in Sec. \ref{geometry}, the first set of properties ---hysteresis,
reciprocity, minimal thickness--- hold generally, for any type of foam or graph.

Convergence and attractors, however, are still conjectural.
They essentially follow from an interplay between determinism of
the process and randomness of the landscape.
Simulations of rectangular foams with periodic boundary conditions in one direction,
infinite in the other direction, confirmed the phenomenon
and gave us quantitative results on the rate of convergence,
its scaling properties and its relation to roughness.

The main biases of our model are that the disorder is confined to one direction
and that the width of the system is finite.
These two aspects, local and global, deserve separate discussions.

\subsection{Anisotropy}

We think that the columnar nature of our model has negligible influence
on our observations; our conclusions hold more generally.
Convergence was probed in the disordered (vertical) direction,
where randomness provides a good imitation of more realistic foams.

Preliminary simulations of topological foams ---generated by
operating a large number of randomly distributed topological transformations
as in \cite{Dubertret,PeStraRi}---
show the same properties as those observed in the rectangular model:
hysteresis, of course, but also, to some extent, convergence of
stratifications, etc.

Notice that convergence is observable and measurable in any type
of foam, not only columnar. This is a significant improvement with respect to
\cite{OgRiv}, where most of the analysis was based on
height $h(x)$, which is rather specific to the columnar model.

\subsection{Boundary conditions}
The extension to foams in other types of spaces is twofold.

As already argued, and shown on an example in concentric geometry,
parallelism, irreversibility and hysteresis, which can be tested
in finite regions, occur quite generally: in planar or 3D
foams, embedded in Euclidean or curved spaces.

As to convergence, it holds unambiguously only in cylindrical foams.
These boundary conditions introduce a definite length-scale into the system.
For quantum gravity, this might be an unbearable hypothesis.
At more common scales, cylindrical geometry is quite frequent.
Condensed matter, zoology or botanic, etc, provide lots of examples
with tubules, channels, stems, stalks, straws,...

When dealing with other boundary conditions, the question of convergence is not
straightforward. There are elementary obstructions
to the onset of a uniform convergence in concentric geometry.
However, convergence may still be true in a weaker sense,
either in the mean over each layer, or restricted to sectors
of prescribed aperture.
All these questions are under current investigations.

%

\begin{thebibliography}{99}
%

\bibitem{weaireRiv}
D. Weaire, N. Rivier,
Contemp. Phys. {\bf 25}, 59 (1984)

\bibitem{weaireHutz}
S. Hutzler, D. Weaire,
{\it The Physics of Foams},
Oxford U.P. (1998)

\bibitem{Aste}
T. Aste, N. Rivier,
J. Phys. A: Math. Gen. {\bf 28}, 1381-1398 (1995)

\bibitem{One} T. Aste, D. Boose, N. Rivier, Phys. Rev. E {\bf 53}, 6181 (1996)\\
N. Rivier, T. Aste, Phil. Trans. R. Soc. London A {\bf 354}, 2055 (1996)\\
T. Aste,
in \textit{Foams and Emulsions} ed. by J.F. Sadoc and N. Rivier, Kluwer (1999), p. 497

\bibitem{Dubertret}
B. Dubertret, N. Rivier, M.A. Peshkin,
J. Phys. A: Math. Gen. {\bf 31}, 879-900 (1998)

\bibitem{Rivier98}
H.M. Ohlenbush, T. Aste, B. Dubertret, N. Rivier,
Eur. Phys. J. B {\bf 2}, 211-220 (1998)\\
H.M. Ohlenbush, N. Rivier, T. Aste, B. Dubertret,
DIMACS Series in Discrete Mathematics and Theoretical Computer Science
{\bf 51}, 279-292 (2000).

\bibitem{OgRiv}
C. Oguey, N. Rivier,
J. Phys. A: Math. Gen. {\bf 34}, 6225 (2001)

\bibitem{Godreche}
C. Godr{\` e}che ed.: \textit{Solids far from Equilibrium}, Cambridge U.P. (1991)

\bibitem{fortesPina}
M. A. Fortes, P. Pina,
{Phil. Mag.} B {\bf 67}, 263 (1993)

\bibitem{OKeeffe}
M. O'Keeffe,
Zeits. f. Kristallogr. {\bf 210}, 905-908 (1995)\\
M. O'Keeffe, S.T. Hyde,
Zeits. f. Kristallogr. {\bf 211}, 73-78 (1996)

\bibitem{Conway}
J.H. Conway, J.A. Sloane,
Proc. R. Soc. Lond. A {\bf 453}, 2369 (1997)

\bibitem{Baake}
M. Baake, U. Grimm,
Zeits. f. Kristallogr. {\bf 212}, 253-256 (1997)

\bibitem{Deza}
G. Binkmann, M. Deza,
Liens {\bf 98-13} (1998), reprinted in Hyperspace {\bf 9}, 28 (2000)\\
(A $n$-corona, as used in this work, is equivalent to the union of layers $j=0, \ldots, n$)

\bibitem{fortes}
M. A. Fortes
J Phys A {\bf 28}, 1055 (1995)

\bibitem{leCaer}
G. Le Caer, J. Phys A {\bf 24} 1307, 4655 (1991)\\
G. Le Caer, R. Delannay, J. Phys. A. {\bf 26}, 3931 (1993);
J. Phys. I France {\bf 5}, 1417 (1995)

\bibitem{schliecker}
G. Schliecker, Phys. Rev. E {\bf 57} R1219 (1998)

\bibitem{frustration}
G. Toulouse, Comm. Phys. {\bf 2}, 115 (1977)\\
J. Villain, J. Phys. C Solid State Phys. {\bf 10}, 1717 (1977)\\
Mattis model: D.C. Mattis, Phys. Letters A {\bf 56}, 421 (1976)\\
Review: N. Rivier, in {\it Geometry in Condensed Matter Physics}, ed. by
J-F. Sadoc, World Scient. 1990, pp. 3-88

\bibitem{KPZ}
M. Kardar, G. Parisi, Y.C. Zhang,
Phys. Rev. Lett. {\bf 56}, 889 (1986)

\bibitem{oguey}
C. Oguey, in preparation...

\bibitem{PeStraRi}
M.A. Peshkin, K.J. Strandburg, N. Rivier,
Phys. Rev. Lett. {\bf 67}, 1803 (1991)

\end{thebibliography}
%


\end{document}